\documentclass[pra,twocolumn,amsfonts,amssymb,amsmath,floatfix,tightenlines,superscriptaddress]{revtex4-1}
\usepackage{amsfonts,amssymb}
\usepackage[subnum]{cases}
\usepackage{mathrsfs}
\usepackage{amsmath}
\usepackage[none]{hyphenat}
\usepackage{graphicx}
\usepackage{hyperref}    % links
\usepackage{footmisc}%\usepackage{cleveref}
\usepackage{bm}          % bold math letters
\usepackage{natbib}
\usepackage{soul} % in order to highlight
\usepackage{color}
\usepackage{longtable}
\usepackage{ textcomp }
\usepackage{verbatim}
\begin{document}

\title{Density waves and jet emission asymmetry in Bose Fireworks}

\author{Han Fu}
%\email{vickeyrobert@uchicago.edu}
\affiliation{James Franck Institute, University of Chicago, Chicago, IL 60637, USA}
\author{Lei Feng}
\affiliation{James Franck Institute, University of Chicago, Chicago, IL 60637, USA}
\affiliation{Enrico Fermi Institute and Department of Physics, University of Chicago, Chicago, IL 60637, USA}
\author{Brandon M. Anderson}
\affiliation{Department of Computer Science, University of Chicago, Chicago, IL 60637, USA}
\author{Logan W. Clark}
\affiliation{James Franck Institute, University of Chicago, Chicago, IL 60637, USA}
\affiliation{Enrico Fermi Institute and Department of Physics, University of Chicago, Chicago, IL 60637, USA}
\author{Jiazhong Hu}
\affiliation{James Franck Institute, University of Chicago, Chicago, IL 60637, USA}
\affiliation{Enrico Fermi Institute and Department of Physics, University of Chicago, Chicago, IL 60637, USA}
\author{Jeffery W. Andrade}
\affiliation{Department of Physics, Harvard University, Cambridge, MA 02138, USA}
\author{Cheng Chin}
\affiliation{James Franck Institute, University of Chicago, Chicago, IL 60637, USA}
\affiliation{Enrico Fermi Institute and Department of Physics, University of Chicago, Chicago, IL 60637, USA}
\author{K. Levin}
\affiliation{James Franck Institute, University of Chicago, Chicago, IL 60637, USA}
\date{\today}

\begin{abstract}
A Bose condensate subject to a periodic modulation of the two-body interactions was recently
observed to emit matter-wave jets resembling ``fireworks" [Nature 551, 356(2017)]. In this paper, combining experiment
with numerical simulation, we demonstrate that these ``Bose fireworks" represent a late stage in a
complex time evolution of the driven condensate. We identify a ``density wave" stage which
precedes jet emission and results from interference of matterwaves. The density
waves self-organize and self-amplify without the breaking of long range
translational symmetry. Importantly, this density wave structure deterministically establishes the template for the subsequent patterns of the emitted jets.
Our simulations, in good agreement with experiment, also address the apparent asymmetry in the jet pattern and show it is fully consistent with momentum conservation.
\end{abstract}

\maketitle

Time-periodic driving, which allows coherent
manipulation of many-body systems, is becoming
an exciting tool in the ultracold atomic gases.
This provides
access to new quantum physics, for example,
topological states, synthetic gauge fields and Mott
transitions
\cite{Eckardt,Jotzu_2014,Monika_chern,Monika_gauge,Zenesini_mott}.
Of particular interest is the rather
unique capability these atomic systems afford into
understanding non-equilibrium many-body dynamics \cite{Vengalattore}.
Also unique to the ultracold gases is the ability, through
the Feshbach resonance, to
periodically modulate atomic interactions \cite{ChengFeshbach}. Recently, this
was implemented by the Chicago group
\cite{Cheng_2017,Logan_gauge}
and
the Rice group
\cite{Hulet_2009, Hulet_2010,Hulet_granulation}
on Bose-Einstein condensates.
In the Chicago experiment, a collective
emission of matter-wave jets resembling fireworks occurs
above a threshold modulation amplitude. The jets
were associated with a form of runaway stimulated
inelastic scattering occurring in the driven
condensate \cite{Cheng_2017}.

\begin{figure}[h]
\includegraphics[width=.485\textwidth]
{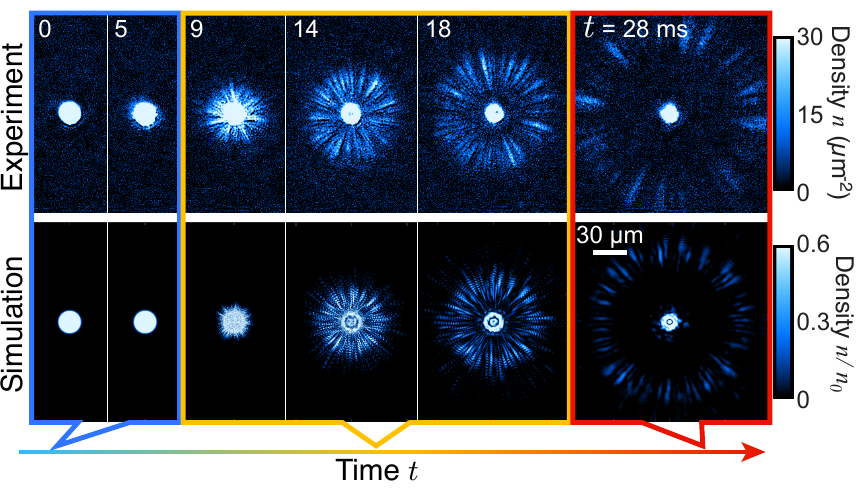}
%{simexpcomp3}
\caption{The real space density distribution
$n({\bf{r}})$ (denoted as $n$) as a
comparison
between simulations and experimental data. In both,
the modulation frequency $\omega/2\pi$ is $2 $ kHz,
the DC and AC interaction energies
respectively are  $U_0n_0 \approx h\times$40~Hz, and
$U_1n_0 \approx h\times$480~Hz,
where $h$ is Planck's constant
(see the main text for detailed definitions).
As a function of modulation time $t$, the system exhibits three phases : density waves in a confined condensate (blue box), near-field emission (orange box) and far-field emission (red box).
}
\label{fig:comp}
\end{figure}

\begin{figure*}
\includegraphics[width=0.8\textwidth]
{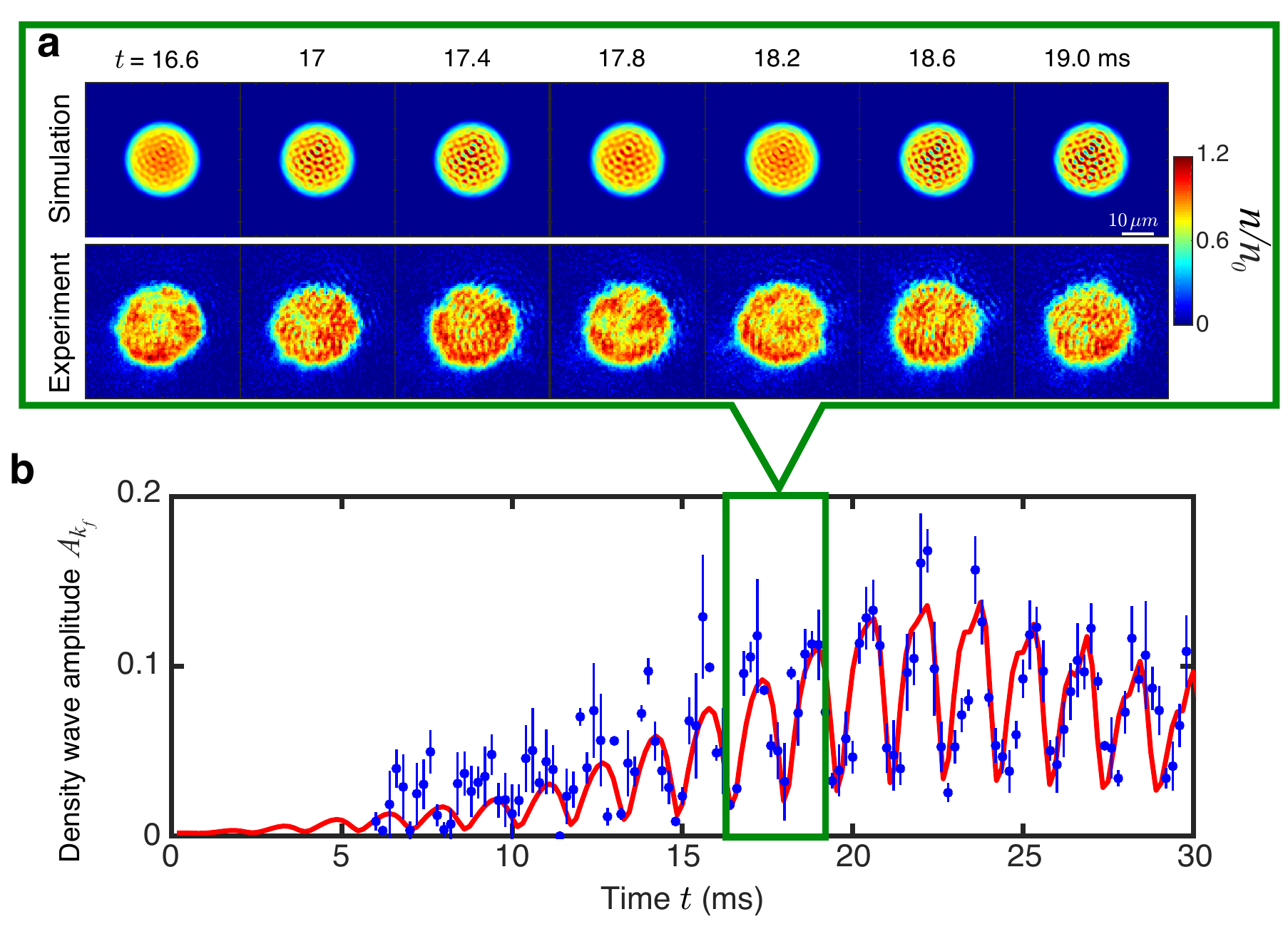}
%{FDW_Oscillation_SimVsExp.pdf}
%{LeiFengBox.pdf}\
\caption
{Experiment and simulation comparison for early-stage density waves (DW)
with $|k| = k_f$. (a) shows the real-space DW oscillations inside the
condensate, comparing theory (top) and experiment (bottom), showing good
qualitative agreement. The experiment exhibits additional static,
long-wavelength density modulations due to trap imperfections. The experimental details are provided in the main text. (b) plots the
amplitude of the density waves in the primary mode comparing simulations
(red solid line) and experiments (blue dots with error bars). In addition
to fast oscillations, both results show consistent observation of an
exponential growth of the envelope until the matter-wave jets are emitted
from the condensate.}
\label{fig:oscillat}
\end{figure*}

\begin{figure}
\includegraphics[width=.48\textwidth]
{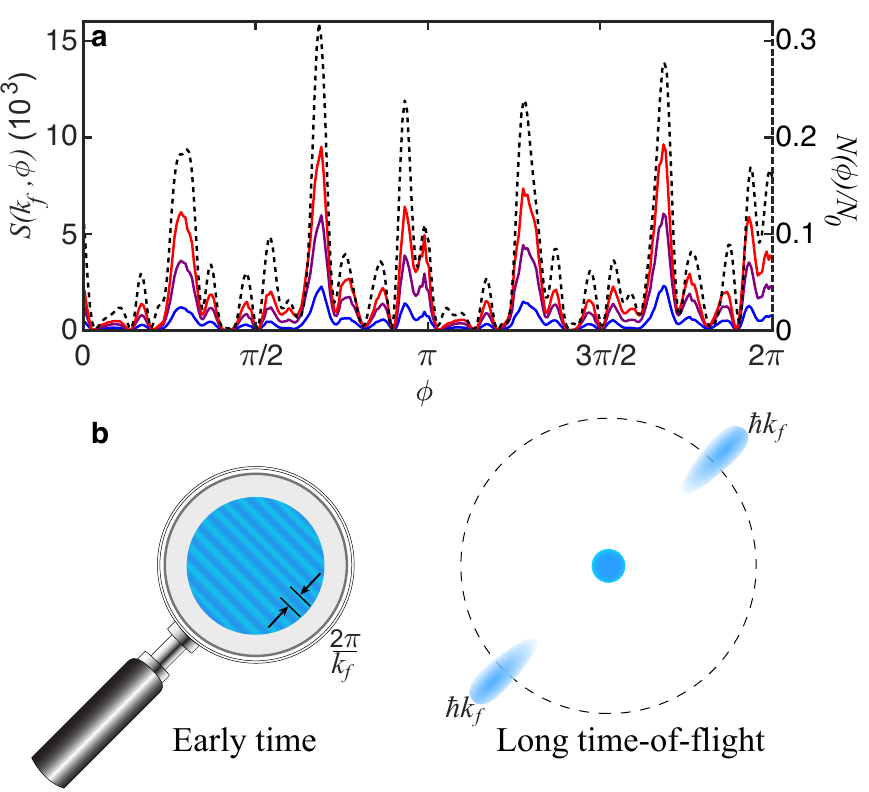}
\caption{Connection between density waves before jet emission and the subsequent matter-wave jet pattern.
(a) shows the azimuthal density structure factor $S(k_f)$ from a single iteration of the GP simulations at resonant wavenumber $k_f$ at $t=10$ (blue), 13 (purple), 15 (red) ms prior to jet emission.
The same shape at different times with growing amplitude is observed and
consistent with the expected amplification process of density waves.
The dashed black curve is the real-space azimuthal population distribution of jets $N(\phi)$ at $t=45$ ms. The scaling factor $N_0$ is the total number of atoms in the system. The alignment of all maxima and minima between $S(k_f,\phi)$ and $N(\phi)$ shows the equivalence between density waves and jets.
(b) schematically shows that the early-time density waves with
wavenumber $k_f$ leads to the emission of counter-propagating jets with
the same wavenumber $k_f$ at long time.}
\label{fig:fingerprint}
\end{figure}

In this paper we use the time-dependent
Gross-Pitaevskii (GP) equation to study the
evolution of the modulated BEC and the emission
of these jets
\cite{Cheng_2017}.
An unbiased or random noise term is introduced initially to model the
fluctuations that seed the jet emission.
%More important are small random
%fluctuations introduced into the condensate
%ground state before the onset of periodic drive.
We show that the simulations capture well the ``fireworks"
dynamics seen in experiments. Moreover, in combination with a new set of experiments, we identify a
previously unobserved stage of the evolution that precedes and underlies
the jet-emission. Immediately after modulation, we
observe that density waves emerge and grow rapidly in the
condensate with quantized wavenumbers determined by the
modulation frequency \cite{Lei}. The density waves arise from the interference between excited matterwaves
and the condensate. The pattern is reminiscent of
Faraday waves in nonlinear
fluids
\cite{Milner,Vinals}
and related to that
predicted for driven atomic gases
\cite{Valcarcel,Kagan, Kevrekidis, Ramaswamy}
as well as observed
in the one-dimensional condensate
\cite{Engels}.

The amplification of these density waves can be considered the
matterwave analog of superradiant scattering in a Bose
condensate
\cite{Ketterle_grating}.
As density waves of wavevector
$\pm{\bf{k}}$
form, they serve as a `grating' within the
condensate to diffract subsequent excitations into the same
counter-propagating modes,
thereby further enhancing
the grating amplitude. This positive feedback leads to the
self-amplification of matter wave excitations.
Such
amplification of the mode population ceases
when the matterwaves leave and are manifested
as the emitted pairs of jets in opposite directions
\cite{Cheng_2017}.

Fig.~1 shows the comparison between
our simulations and experiment, and serves as a calibration throughout
this paper. Importantly,
it suggests, by the three different colorations,
that three distinct regimes of the Bose
fireworks can be identified: the early density wave (DW)
regime, the initial emergence of jets (called the ``near-field emission")
and the well established jet emission regime
(called the ``far-field emission"). In the near field stage
the excitation modes begin to leave the condensate while still
maintaining a substantial overlap with each other. After
sufficiently long time (in the far field), the matter wave jets
are well separated. Here the emission pattern shows
that well resolved and fully distinct momentum modes
are populated.

We begin with the theoretical and
experimental investigation of the early-stage
density waves.
Fig.~2 presents
the experimental observation and theoretical confirmation
for the emergence of density waves.
In experiment we start with a Bose condensate of
$4\times10^4$
cesium atoms prepared in a uniform disk-shaped
trap with a radius of $ 13\, \mu$m. (See Ref.~\onlinecite{Cheng_2017}
for experimental details.) The trap has a potential barrier of
height
$h\times 200$~Hz in the horizontal direction ($h$ is the Planck constant)
and is harmonic vertically with a
frequency 220~Hz. By modulating the magnetic field
near a Feshbach resonance, we oscillate the scattering
length as
$a(t)=a_{\rm dc}+a_{\rm ac}\sin (\omega t)$
with a small offset
$a_{\rm dc}=4 a_0$
and large amplitude
$a_{\rm ac}= 40 a_0$
at frequency
$\omega/ 2\pi =620$~Hz,
where $a_0$ is the
Bohr radius.

After oscillating the interaction for time $t$, we perform \textit{in situ} imaging and observe density waves forming within the condensate prior to jet emission. Shown in Fig.~2 (a) are snapshots of the condensate density distribution $n(\mathbf{r})$ and theoretical simulation. To be more quantitative,
we extract the density wave amplitude $A_{k_f}$ from the Fourier transform of the condensate $\tilde{n}(\mathbf{k})=(2\pi)^{-1}\int d \mathbf{r}e^{-i\mathbf{k}\cdot\mathbf{r}}n(\mathbf{r})$, and plot $A_{k_f} =n_0^{-1}\int_{|\mathbf{k}|=k_f}d{\bf{k}}\,|\tilde{n}(\mathbf{k})|$ as a function of time, see Fig.~2 (b). Here $k_f= \sqrt{m \omega/\hbar}$ is the wavenumber of the
density waves which is determined by the parametric resonance condition; $n_0$ is the average density of the static condensate prior to interaction oscillation; $m$ is the boson mass, and $\hbar$ is the reduced Planck constant. Interestingly, this density wave amplitude exhibits a fast oscillation under a slowly growing envelope.

%\vskip1mm \noindent \textbf{Overview of Theory-}
Our theoretical approach is based on
a dynamical GP equation:
\begin{equation}\label{eq:GP}
\begin{aligned}
i\hbar\frac{\partial\psi}{\partial t}=&\left[-\frac{\hbar^2}{2m}\nabla^2+V({\bf{r}})+U_0|\psi|^2 -\mu\right]\psi \\
&+U_1\sin(\omega t)|\psi|^2\psi,
\end{aligned}
\end{equation}
where $\psi$ is the wavefunction, and $\mu=U_0n_0$ the chemical potential of the static condensate;
$V({\bf{r}})$ is the external trap potential, and ${\bf{r}} = (x,y)$ is a two-dimensional (2D) spatial coordinate (with origin
at the trap center).
In addition $U_0=4\pi\hbar^2a_{\rm dc}/m$ and $U_1=4\pi\hbar^2a_{\rm ac}/m$ are
the DC and AC interaction strengths, respectively;
here $a_{\rm dc}>0$ is the background offset, $a_{\rm ac}$ is the
amplitude of the
AC scattering length.
At short times, the condensate is weakly excited and the
wavefunction can be linearized \cite{Valcarcel, Kagan}
\begin{equation}\label{eq:expansion}
\psi=\psi_0\left[1+\nu({\bf{r}}, t)\right],
\end{equation}
where $\psi_0=\sqrt{n_0}\exp\left[iU_1n_0\cos(\omega t)/\hbar\omega\right]$ is the wavefunction of a uniform BEC, and $U_0$ has been
absorbed through the parametrization in Eq.~(\ref{eq:GP}).
Since the characteristic DW length scales are much smaller than the trap size,
we ignore trap effects in our analytical approach.
In the plane wave basis we write
$\nu({\bf{r}},t)=\left[\xi(t)+i\zeta(t)\right]\cos(\bf{k\cdot r}+\varphi)$ with both $\xi(t)$ and $\zeta(t)$ real and $\varphi$ a random phase. Since $|\nu|\ll1$, $\xi$ satisfies the
Mathieu equation for parametric resonances:
\begin{equation}\label{eq:Mathieu}
\frac{\partial^2 \xi}{\partial t^2}+\Omega^2\left[1+\alpha \sin(\omega t)\right]\xi=0,
\end{equation}
and $\zeta$ satisfies the same equation with an extra term
$-\alpha\omega\cos(\omega t)\frac{\partial\zeta}{\partial t}$ on the left hand side.
Here we keep only leading terms in $\alpha$; $\Omega^2=\hbar^2k^4/4m^2+U_0n_0k^2/m$, and $\alpha=U_1n_0k^2/m\Omega^2$.

The solution of Eq.~(\ref{eq:Mathieu})
is $\xi(t)
\approx A_+ \cos(\omega t/2+\vartheta_+)\exp(\lambda_+ t)+A_-\sin(\omega t/2+\vartheta_-)\exp(\lambda_- t)$.
Here $A_{\pm}$ are numerical coefficients, and the exponents are
\begin{equation}\label{Eq:densityexp}
\lambda_{\pm}=\pm\sqrt{\frac{\alpha^2\Omega^2}{16}-\left(\Omega-\frac{\omega}{2}\right)^2}.
\end{equation}
The solution exhibits both subharmonic oscillations with half the driving frequency $\omega$ and an exponential envelope growth (via $\lambda_+$). For $U_0\approx 0$ as in
experiments, the resonance with maximal $\lambda_+$ occurs at $k=k_f$. At this point, $\vartheta_\pm\approx0$, and $\zeta(t)\approx -A_+ \sin(\omega t/2)\exp(\lambda_+ t)+A_-\cos(\omega t/2)\exp(\lambda_- t)$.

The interference between the uniform background and the excitations then gives the density $n({\bf{r}})=n_0|1+\nu({\bf{r}},t)|^2\approx n_0\left[1+2\xi(t)\cos({\bf{k}\cdot{\bf{r}}}+\varphi)\right]$, leading to the density waves of exponentially growing envelope that we
report here.
To provide the full dynamical evolution and to include
trap effects, we next appeal to the more complete numerical simulations of the GP equation.
%CAN WE REMOVE THIS HERE The density waves result from the interference between the condensate
%and plane waves. After the excitations escape the condensate, the observed emission structure at short times reflects the overlapping excitations in the near field; these
%are well separated in the far field.
%

Our simulations are 2D and incorporate a ring trap
with inner and outer radii $R_{\rm in}$ and $R_{\rm out}$, respectively. We choose
$V({\bf{r}}) = V_0$ for $R_{\rm in}<r<R_{\rm out}$
and zero elsewhere. $V_0$ is taken to be compatible with experiment,
$R_{\rm in}$ is taken to be the condensate radius, and, as in experiment
\cite{Cheng_2017}, $R_{\rm out} \approx 1.5 R_{\rm in}$.
We use a CUDA-based GP equation solver \cite{Kathy,ourKZ}, implemented on graphic processing units (GPU). More specifically, we adopt a split-step algorithm with a spectral technique in momentum space to evolve the condensate wavefunction forward in time.
At $t>0$ we introduce a periodic oscillation of the two-body interaction term.
Important are initial small (of order $1/\sqrt{N_0}$, where $N_0$ is the total number
of particles) random
fluctuations introduced into the condensate
ground state before the onset of periodic drive \footnote{This fluctuation term $\psi_s = \varepsilon_r + i \varepsilon_i$ is added to the initial ground state
wave function $\psi_0$. Here we chose the random variables $\varepsilon_r$  and $ \varepsilon_i$ to have a Gaussian probability density function centered around zero with standard deviation $\sigma = 0.1|\psi_0|$. }.

\begin{figure}
\includegraphics[width=.5\textwidth]
{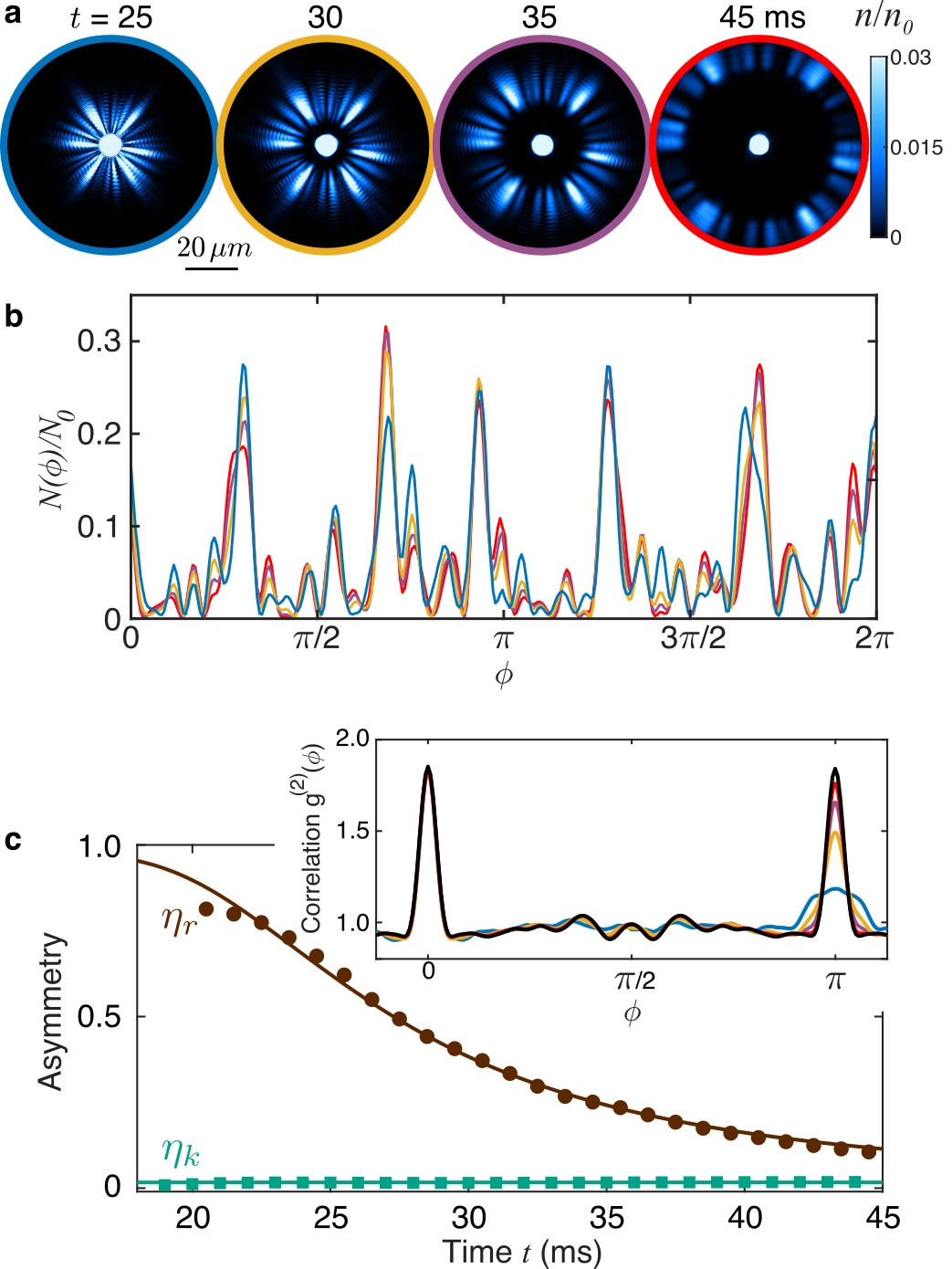}
%{Fig4_new04.png}
%{Fig1_new01.pdf}
\caption
{Time evolution and correlations of the emitted jets. (a) shows the calculated jet emission pattern evolving
from the near- to far-field regimes. The calculation is based on identical initial noise seeding. (b) shows the real space azimuthal population of the four images
in (a), identified by the same color. Note that the $t = 45 $ ms far-field curve is equivalent to that shown
dashed in
Fig.~\ref{fig:fingerprint}~(a). Here unlike in Fig.~\ref{fig:fingerprint},
the peaks and valleys are slightly displaced
with time. Panel (c) probes the emission asymmetry in real space $\eta_r = g^{(2)}(\pi) - g^{(2)}(0)$ (brown circles) and the momentum-space analogue $\eta_k$ (green squares). The
main figure shows that the $(0,\pi)$ asymmetry
is always absent in momentum space ($\eta_k$ is strictly zero within numerical precision) so that momentum is conserved.
In real space, using Panel (b) we find that
this $(0,\pi)$ asymmetry decreases with increasing time.
The inset indicates
the correlation function $g^{(2)}(\phi)$ at the same 4 times in (a),
along with an
early time momentum correlation function at $t=20$ ms (black curve). % that also corresponds to long TOF. %This represents the equivalence of the density wave CHECK and momentum space population.
Again,
inversion 0-$\pi$ symmetry is broken at short times, but recovers after long time-of-flight, and is fully preserved in momentum space. The solid line (brown) in (c) is an
analytical fit to $\eta_r$ \cite{Supple}.}
\label{fig:pipeak}
\end{figure}

It should be noted that
the exponents in Eq.~(\ref{Eq:densityexp})
coincide with those derived in Ref.~\onlinecite{Cheng_2017} for
the matter-wave jets.
This suggests that the two forms of excitations may be manifestations of the same physics.
We probe this hypothesis in Fig.~\ref{fig:fingerprint} which contains results from our full GP simulations.
%This figure shows in more detail evidence for the equivalence between density waves and jets.
%
Indeed, Fig.~\ref{fig:fingerprint} provides strong simulation evidence that the density waves are
necessary precursors to the jets and that they
establish the template for the subsequent jet emission pattern.
We do this by demonstrating that the structure factor with
fixed extrema
(established by the DW pattern at the onset of shaking)
is precisely equivalent
to the real-space emitted jet population
$N(\phi)$
at long times, through time of flight.

The structure factor is
defined by
$S(k_f,\phi) =  N_0^{-1}\int k dk|\tilde{n}({\bf{k}})|^2$
(where the magnitude
and phase of the wavevector arguments are $|{\bf{k}}|=k\approx k_f$ and $\tan\phi=k_x/k_y$).
Note from
Fig.~\ref{fig:fingerprint} (a)
that the structure factor
contains random peaks and valleys as determined by the initial random seed which emulates the fluctuations of real experiments.
While these are
established
at the onset of shaking, with increasing time the only
change is an exponential growth of the peak amplitudes.

The dashed black line plotted in Fig.~3 (a) is the real-space azimuthal distribution
for the jet population $N(\phi)=\int_{{\bf{r}}=(\hbar t/m){\bf{k}}}rdr\, n({\bf{r}})$, at long times. Importantly,
the angular distribution shows the equivalence between $S(k_f,\phi)$ and $N(\phi)$. This
underlies our claim that density waves and jets are deterministically
correlated. These results are summarized in Fig.~\ref{fig:fingerprint}~(b). This
presents a schematic plot linking the momentum space
spectrum of the DW
%(through the structure factor)
and
the population of jets with the same wavevector $\pm{\bf{k}}$ after long time of flight.

%These results are summarized in Fig.~\ref{fig:fingerprint} (b) which presents a schematic plot relating to the concepts behind Fig.~3. The figure on the left emphasizes that the structure factor, plotted above, is to be associated with the early-time density waves in the condensate. The figure on the right makes the point that the real-space population $N(\phi)$ measures the emitted jet distribution through time of flight at long times.

Having established the equivalence between the far-field jets and the
initial density waves, one might expect that in the near field when jets
are first emitted from the condensate, the same azimuthal distribution
profile in real space would be retained. Our simulations, however, show that
this is not the case. In Fig.~\ref{fig:pipeak}~(a) and (b), a clear modification of the distribution shape
with varying time is seen and is accompanied by an ``inversion symmetry
breaking" (in the near field). This is associated with the observation
(first reported experimentally \cite{Cheng_2017}) of
an asymmetric
two-particle correlation function $g^{(2)}(\phi)$ of the jet emission
pattern, i.e., $g^{(2)}(\pi)\neq g^{(2)}(0)$. It has been attributed to
momentum non-conservation \cite{Hui} or alternatively a ``di-jet
acollinearity" observed, for example, in quark-gluon plasmas
\cite{Arratia}.

Here we propose and provide strong numerical support for a different scenario
which is well substantiated
by
the detailed numerics which are summarized in Fig.~\ref{fig:pipeak}~(c), along
with analytical arguments in the supplementary material \cite{Supple}.
To quantify this inversion asymmetry,
we introduce a parameter
$$\eta_r = \frac{\left\langle\left[N(\theta)-N(\theta+\pi)\right]^2\right\rangle}{2\left\langle N(\theta)\right\rangle^2}=g^{(2)}(0)-g^{(2)}(\pi)$$ for real space (and its analogue,
$\eta_k$ in momentum space \cite{Supple}), where $\left\langle\ldots\right\rangle$
corresponds to averaging over angles $\theta$ and ensembles. Fig.~\ref{fig:pipeak}~(c)
plots the the asymmetry functions,
$\eta_{r,k}$, in real- and momentum-space, together with the corresponding correlation function $g^{(2)}(\phi)$ shown in the inset. The spatial asymmetry $\eta_{r}$ decreases from a finite value to zero when going from the near to far field. This
indicates that the inversion symmetry is recovered at large times. The momentum-space asymmetry $\eta_k$, interestingly, remains strictly zero
independent of time, showing clearly that momentum conservation is obeyed at all times.

We attribute this asymmetry to the fact
that, in the near field, excitations of different wavevectors substantially
overlap with each other. The resulting pattern is derived from interference between
these overlapping modes, which have uncorrelated random phases.
%This leads
%to a density $n({\bf{r}})\approx n_0|\sum_{{\bf{k}}} \nu_{\bf{k}}({\bf{r}},t)|^2=n_0\sum_{\bf{k},\bf{k'}}\nu_{\bf{k}}({\bf{r}},t)\nu^*_{\bf{k}'}({\bf{r}},t)$.
Thus, when measuring the population at angles $\theta$ and $\theta+\pi$, the symmetry between the relevant counter-propagating pair $\pm\bf{k}$ ($\tan\theta=k_y/k_x$), is masked by interference from other uncorrelated modes.
By contrast, in the far field, different modes are well separated so that each
jet now represents a single mode. Here momentum conservation is more
apparent and inversion symmetry in real space is recovered.
We emphasize this physical picture \cite{Supple}
is different from other scenarios \cite{Arratia,Hui} in the literature.

\vskip2mm
\noindent
\textit{Conclusions.--}
In this paper we have investigated the jet emission process induced by a periodic drive of the two-body interactions. Through a combination of simulations of the Gross-Pitaevskii equation and experiments, we demonstrated that the jet structure is imprinted in the early stages of an excited condensate, through density waves. These density waves
set up an effective self consistently produced grating which, through feed-back effects
resonantly amplifies their pattern \cite{Ketterle_grating}.
What is different from the literature \cite{Ostermann,Valcarcel} is that the grating here is disordered, but, nevertheless,
the amplification process proceeds and ultimately leads to the ejection of jets or `` Bose fireworks".

Observing the actual density waves in experiments, as reported in the present paper, was key to confirming this picture. Also critical to this analysis is the
demonstrated capability of the GP
simulations to successfully address experiments involving
this stimulated emission over widely
varying time, space and momentum coordinates.
Our simulations have provided predictive capabilities as well as
the ability
to establish the important underlying principles (such as momentum conservation) of this
broad scope of experimental matter-wave jet observations.

\phantom{}

{\em Acknowledgments.}

We are
grateful to Tom Witten for helpful discussions and Igor Aronson and
Andreas Glatz for the numerical GP code. L. F. acknowledges support from an
MRSEC-funded Graduate Research Fellowship.  L. W.
C. was supported by a Grainger
Graduate Fellowship. This work was primarily supported by the University of
Chicago Materials Research Science and Engineering Center, which is funded
by the National Science Foundation under award number DMR-1420709. We also
acknowledge support from NSF Grant No. PHY-1511696 and the Army Research
Office-Multidisciplinary Research Initiative under grant W911NF-14-1-0003.

\pagebreak
\widetext
\begin{center}
\textbf{\large Supplement: Density waves and jet emission asymmetry in Bose fireworks}
\end{center}
\setcounter{equation}{0}
\setcounter{figure}{0}
\setcounter{table}{0}
\renewcommand{\theequation}{S\arabic{equation}}
\renewcommand{\thefigure}{S\arabic{figure}}
\renewcommand{\bibnumfmt}[1]{[S#1]}
\renewcommand{\citenumfont}[1]{S#1}
\makeatother

\noindent
\section{Contrasting behavior of jets from the real- and momentum-space perspectives}
\vskip2mm

In this supplement we address the contrasting behavior between real- and
momentum-space behavior of the emitted jets as a function of time.
At issue here is the possibility of momentum non-conservation during the jet emission process which some authors \cite{Arratia_supp,Hui_supp} have
associated with the $(0,\pi$) asymmetry found in the
two-particle correlation function.
Our simulations have shown momentum is always conserved
even with this asymmetry,
and we present additional support for this important claim in this
supplementary material.

We focus on
the two-particle correlation function $g^{(2)}(\phi)$
in real space and the relationship between its
behavior at $\phi=0$ and $\phi=\pi$
(where $\phi$ is an azimuthal angle). An asymmetry is
experimentally \cite{Cheng_2017_supp} seen at short times. The reason for this asymmetry
in real space is discussed here and we show as well that
in momentum space this asymmetry is absent at all times.

We define the
two-particle correlation function $g^{(2)}$ in real space as
\begin{equation}\label{eq:correlation}
\begin{aligned}
g^{(2)}(\phi)=&\frac{\left\langle n(\theta)\,n(\theta+\phi)\right\rangle}{\left\langle n(\theta)\right\rangle^{\,2}},
\end{aligned}
\end{equation}
where $n(\theta)$ is the particle population at azimuthal angle $\theta$. %The corresponding $g^{(2)}$ in momentum space has the same expression except for that n(\theta) is replaced by the $k$-space particle population $n_{k_f}(\theta)$ where $n_{k_f}(\theta)=n({\bf{k}})$ at $|{\bf{k}}|=k_f,\, \tan\theta=k_y/k_x$. Here
In real space, this corresponds to density $n({\bf{r}})$ at the position ${\bf{r}}=r(\cos\theta,\sin\theta)$. We can also define the momentum-space analogue of $g^{(2)}(\phi)$, where $n(\theta)$ refers to $n({\bf{k}})$ at ${\bf{k}}=k_f(\cos\theta,\sin\theta)$. Here $r=v_ft$ and $v_f=\hbar k_f/m$ is the jet velocity. $n({\bf{r}})=|\psi({\bf{r}})|^2$, $n({\bf{k}})=|\psi({\bf{k}})|^2$, and $\psi({\bf{k}})=(2\pi)^{-1}\int d \mathbf{r}e^{-i\mathbf{k}\cdot\mathbf{r}}\psi(\mathbf{r})$ is the Fourier transform of
the wave function $\psi({\bf{r}})$. $\langle\ldots\rangle$ refers to an average over different angles $\theta$ and ensembles.

\begin{figure}[h]
\includegraphics[width=0.6\textwidth]
{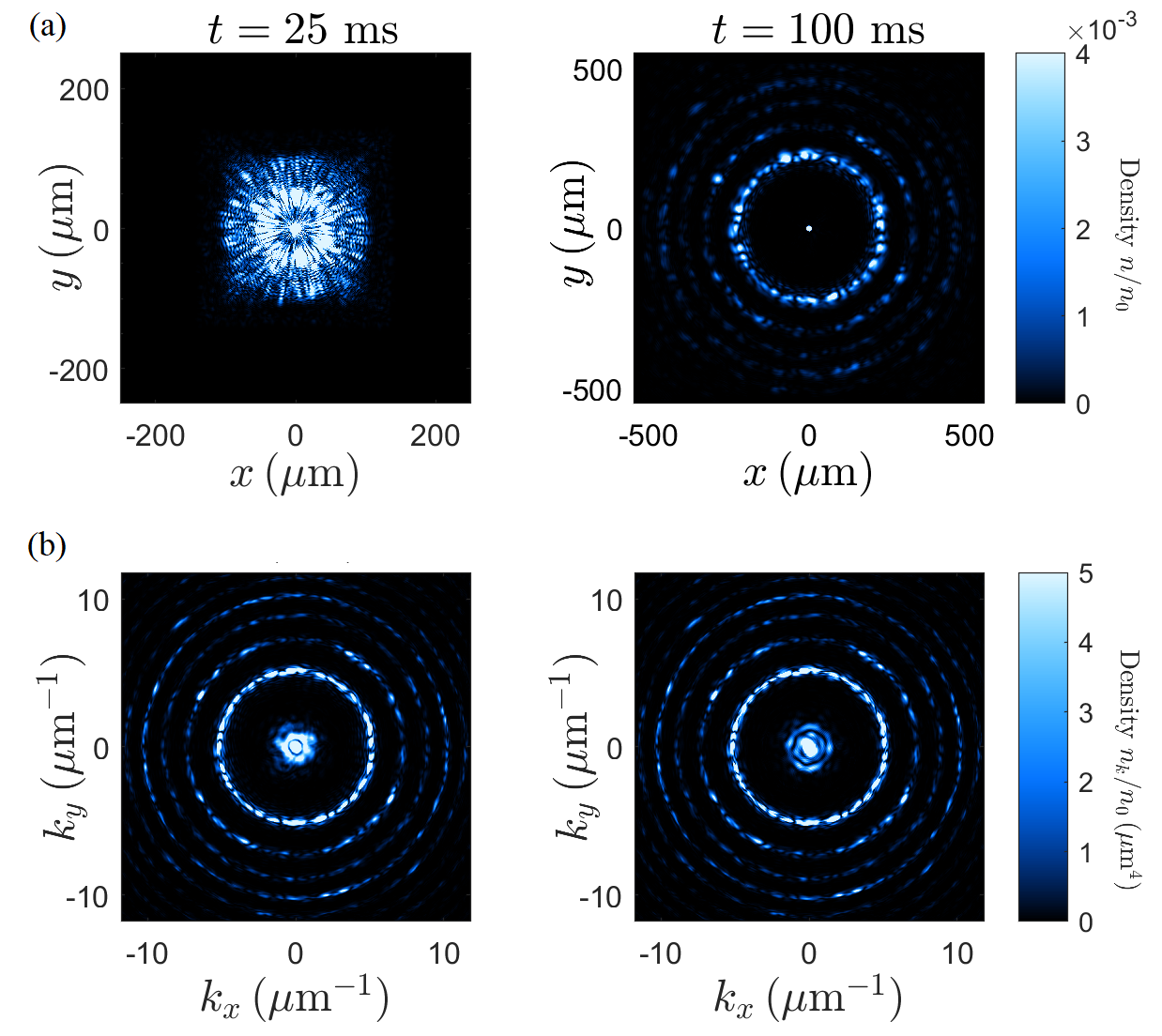}
\caption{Particle distributions at early and late times. (a) Real-space distribution $n({\bf{r}})$ (denoted by $n$). The TOF distribution changes
substantially with growing time and approximately reproduces the $k$-space
results at large $t$. (b) Momentum-space distribution $n({\bf{k}})$ (denoted by $n_k$). A primary ring (along with weak
secondary rings representing higher harmonics) shows little variation with time.  The left column shows the early-time behavior ($t=25$ ms) when jets have just emerged. The right column indicates the late-time distribution ($t=100$ ms) when jets are far away from the trap.}
\label{fig:nearfar}
\end{figure}

\begin{figure*}[h]
\includegraphics[width=.85\textwidth,clip]
{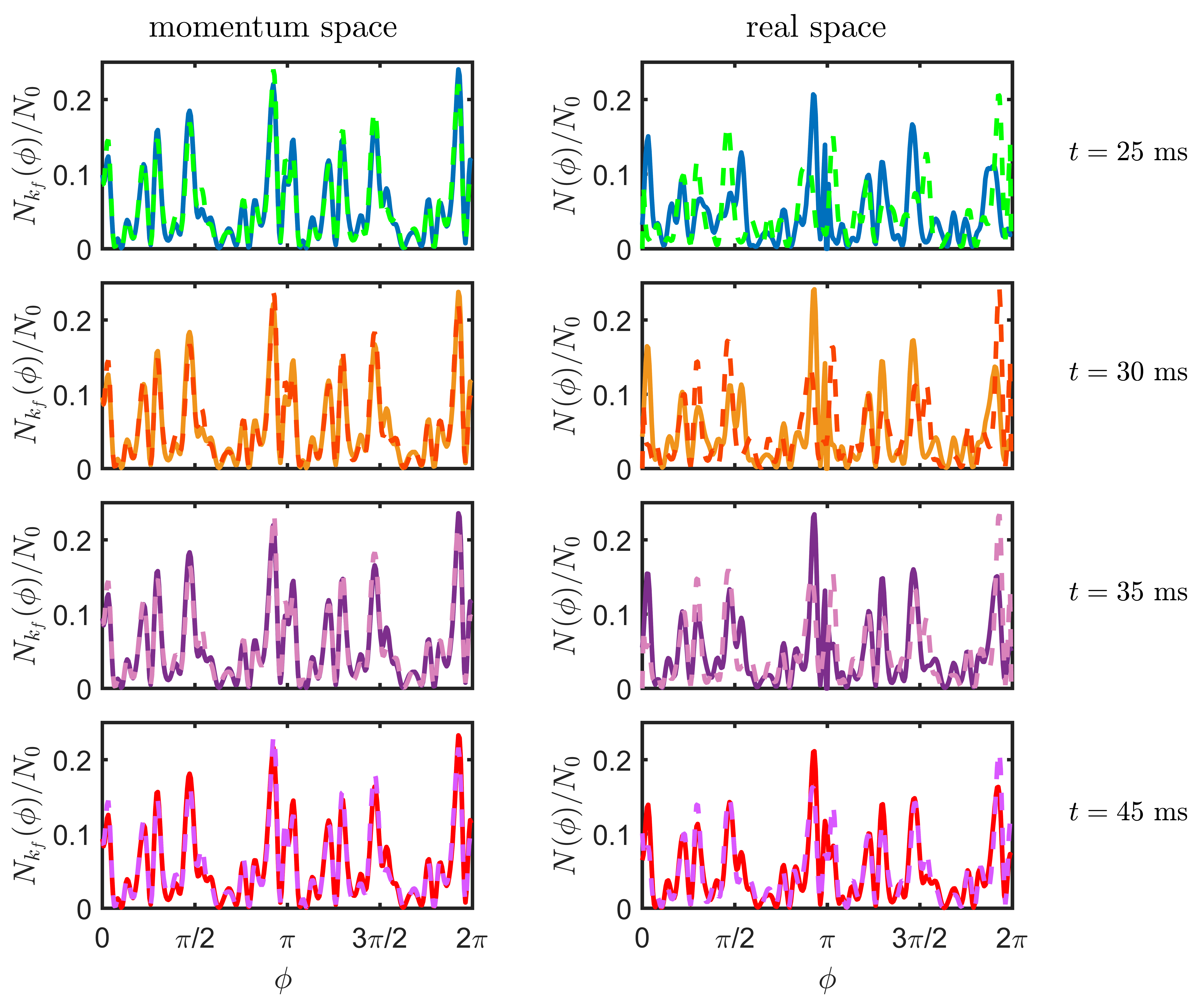}
\caption{momentum- and real-space $\pi$-shifted plots. The left column represents the $k$-space azimuthal number distribution $N_{k_f}(\phi)$ (solid line) superposed with its $\pi$-shifted curve (dashed line) $N_{k_f}'(\phi)=N_{k_f}(\phi+\pi)$. The right column corresponds to the real-space population $N(\phi)$ (solid line) with its $\pi$-shifted curve (dashed line) $N'(\phi)=N(\phi+\pi)$. Different colors correspond to different times: $t=25$ ms (blue), $t=30$ ms (yellow), $t=35$ ms (purple), and $t=45$ ms (red). Peaks of $N_{k_f}(\phi)$ and $N'_{k_f}(\phi)$ are well aligned in both short and long times, while $N(\phi)$ and $N'(\phi)$ are misaligned at small $t$ and well aligned at large $t$.}
\label{fig:kfing}
\end{figure*}

Fig.~\ref{fig:nearfar} presents the particle density distribution in
real and momentum space in (a) and (b), respectively.
The left column corresponds to early times where the jets are just
emerging while the right column is for long times. In momentum space
a primary ring (and weak secondary rings) are visible and one sees
very little time dependence; this is in contrast to the real-space plots. Nevertheless
at longer times it is evident that the real-space time of flight (TOF)
appears to reproduce the $k$-space distribution of particles. This
is expected and, as a corollary implies that at large $t$, the peaks in
the real-space $g^{(2)}$ at $0,\,\pi$ become symmetric.
Indeed this is observed in
Fig.~\ref{fig:pipeak} (c) of the main text.

To gain further insight, in
Fig.~\ref{fig:kfing} we plot the
momentum- (left column) and real-space (right column) azimuthal number distributions.
These correspond, respectively to
$N_{k_f}(\phi)$ and
$N(\phi)$, where $N_{k_f}(\phi)=\int_{{\bf{k}}'={\bf{k}}}kdk\, n({\bf{k}}')$ and ${\bf{k}}=k_f(\cos\phi,\sin\phi)$.
Plotted as solid lines in each row are distributions for the
same 4 indicated times as in Fig.~\ref{fig:pipeak} of the main text.
The dashed lines correspond to the same plot with each angle
shifted by $\pi$.
This corresponds to the shifted distribution $N_{k_f}'(\phi)=N_{k_f}(\phi+\pi)$ and $N'(\phi)=N(\phi+\pi)$, respectively.

In the momentum-space plot of Fig.~\ref{fig:kfing} one can see an essentially exact
coincidence of the solid and dashed curves showing the full
symmetry between excitations with opposite momentum. This
occurs for all times and is a manifestation of momentum conservation throughout.
This behavior should
be contrasted with plots of the real-space number distribution in
Fig.~\ref{fig:kfing}
where one can see quite generally,
that the peaks of $N(\phi)$ and in the shifted distribution $N'(\phi)=N(\phi+\pi)$ are misaligned.
Importantly, only at the latest times (in the far field)
is there a complete overlap of the curves which translates into
a $(0,\pi)$ symmetry for the correlation function $g^{(2)}(\phi)$.

\section{Simple analytic model of near-field correlation functions}

To understand the origin of this symmetry breakdown in real space for near fields, we appeal to a simple analytic model.
This model builds on our understanding
that the phase remains correlated only within the same excitation mode.
We demonstrate this later.

\begin{figure}[h]
\includegraphics[width=.45\textwidth]{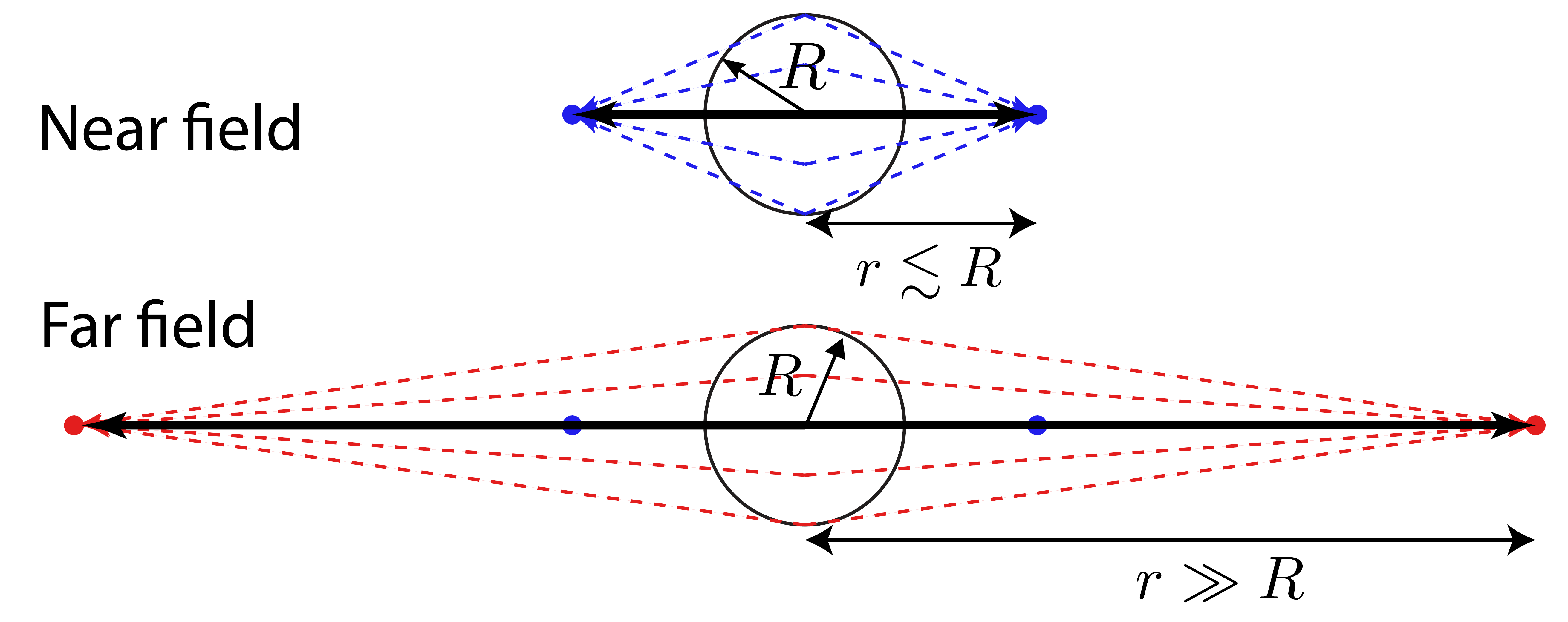}
\caption{Schematic of near- and far-field TOF. In both near and far fields, the only correlated modes are the counter-propagating pair. In the far field where the distance between jets and trap center (called $r$) is much bigger than trap size $R$, fewer modes appear
to overlap as seen from the measurement point.}
\label{fig:g2}
\end{figure}

\begin{figure*}
\includegraphics[width=\textwidth]
{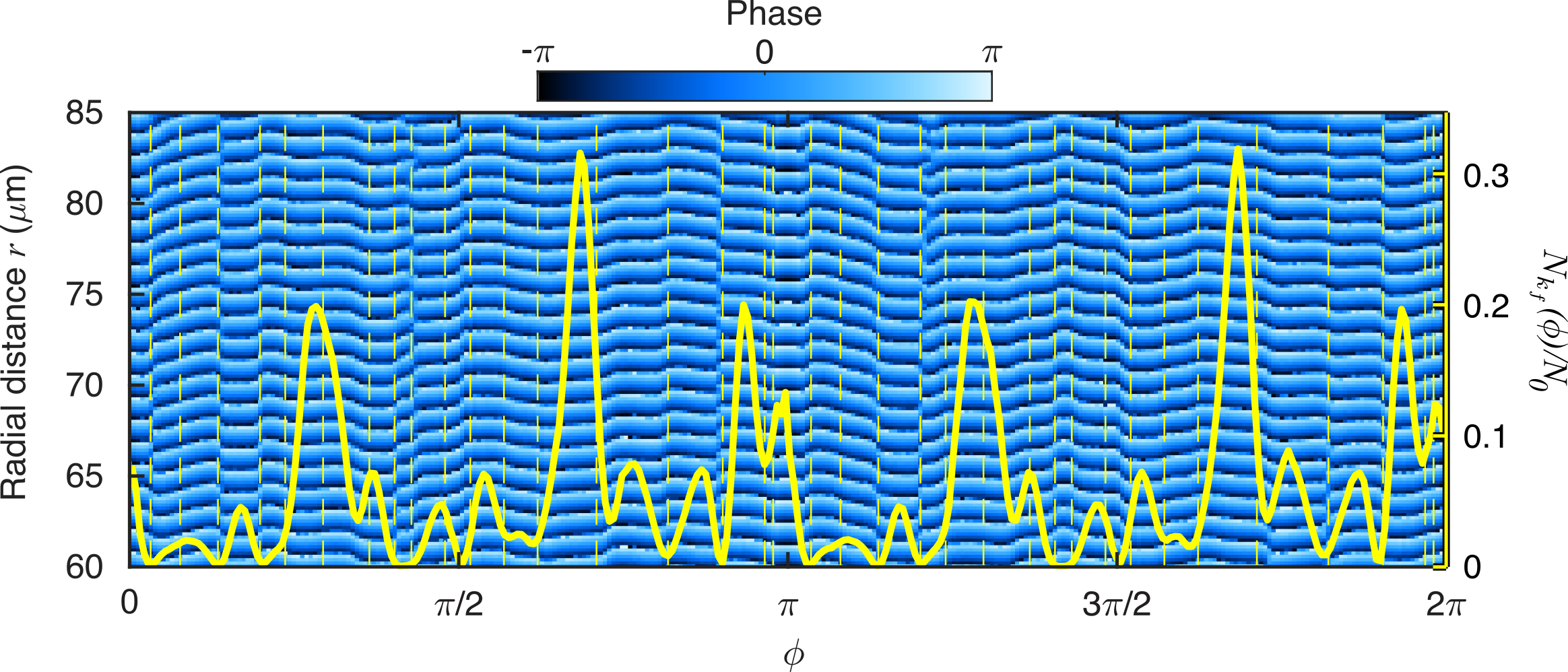}
%{phaseLei}\
\caption{Intra-jet phase coherence in far field. Comparison of GP
wave-function phase and
momentum distribution of jets
as a function of the azimuthal angle $\phi$ for the relevant radial range where the jets are located.
Dashed lines have been inserted to
mark phase discontinuities.
%In the far field, each jet represents a single mode.
The yellow curve is the momentum distribution which represents
the far field spatial configuration of the jets. Within each jet there is phase coherence. However,
coherence typically ends at the boundary between different jets (modes) where a ``phase slip"
can be seen. It should be noted that
the applicability of the ``edge finding" algorithm used here leads to occasional
errors when the phase slips are found to be particularly small.
}
%The mismatch of jet boundary and phase slips at certain places is likely due to imprecision of manual insertion.
\label{fig:phase}
\end{figure*}

As shown in Fig.~\ref{fig:g2}, in both
the near-field and far-field situations, the only perfectly correlated
pair of modes is the counter-propagating one going through the center of the condensate (marked as thick black lines with arrow heads). %This is consistent with observations that other pairs of jets are phase-uncorrelated (see Fig.~\ref{fig:phase}).
As a result, the wave function at $\bf{r}$ and $-{\bf{r}}$ can be approximated as
\begin{equation}\label{eq:jets}
\begin{aligned}
\psi({\bf{r}})&=\sqrt{n_1}e^{i\varphi_{1}}+\sum_{j=2}^{m}\sqrt{n_{j}}e^{i\varphi_{j}}\\
\psi(-{\bf{r}})&=\sqrt{n_1}e^{-i\varphi_{1}}+\sum_{j=2}^{m}\sqrt{n_{j}^{\prime}}e^{-i\varphi_{j}^{\prime}},
\end{aligned}
\end{equation}
where $n_{1}$ is the occupation of the perfectly correlated modes
while $n_{j}$ and $n_{j}^{\prime}$ are the occupations of modes propagating
in different directions.
%$\varphi_i$ and $\varphi_{j\neq i}$, $\varphi_i'$ and $\varphi_{j\neq i}'$, or
Note $\varphi_i$ and $\varphi_{j}'$ are uncorrelated.

We now want to count the number of overlapping
modes $m$ at the measurement point $r=v_f t$ forming the jet.
Entering into the count is
$\Delta\phi/\delta\theta=\arctan(R/r)/\delta\theta=\arctan(R/v_ft)/\delta\theta$
with $\delta\theta\sim 1/k_fR$ representing the angular half width of the jets.
This provides a reasonable estimate of $m$ in the near field. However, this is
inadequate in the far field because
it approaches zero, rather than the expected 1. Thus a more
appropriate, phenomenological estimate would be to add the near- and far-field estimates in quadrature
\begin{equation}
m=\sqrt{1+\left(\frac{\Delta\phi}{\delta\theta}\right)^2}.
\end{equation}

Using Eq. \eqref{eq:jets}, the correlation function can then be written as
\begin{equation}\label{eq:correlation}
\begin{aligned}
g^{(2)}(0) & =\frac{\left\langle\psi^*({\bf{r}})\psi({\bf{r}})\psi^*({\bf{r}})\psi({\bf{r}})\right\rangle}{\left\langle\psi^*({\bf{r}})\psi({\bf{r}})\right\rangle\left\langle\psi^*({\bf{r}})\psi({\bf{r}})\right\rangle}\\
g^{(2)}(\pi) & =\frac{\left\langle\psi^*({\bf{r}})\psi({\bf{r}})\psi^*(-{\bf{r}})\psi(-{\bf{r}})\right\rangle}{\left\langle\psi^*({\bf{r}})\psi({\bf{r}})\right\rangle\left\langle\psi^*(-{\bf{r}})\psi(-{\bf{r}})\right\rangle}.\\
\end{aligned}
\end{equation}
And we have
\begin{equation}
\begin{aligned}
&\left\langle \psi^*({\bf{r}})\psi({\bf{r}})\right\rangle =\left\langle\psi^*(-{\bf{r}})\psi(-{\bf{r}})\right\rangle \\
=&\sum_{j=1}^{m}\left\langle n_{j}\right\rangle+\left\langle\sum_{i\neq j}\sqrt{n_{i}n_{j}}\cos(\varphi_{i}-\varphi_{j})\right\rangle\\
=&\,m\left\langle n\right\rangle,
\end{aligned}
\end{equation}
where $\left\langle n\right\rangle$ is the average density associated with a jet. The same-site
correlation function is
\begin{equation}
\begin{aligned}
&\left\langle \psi^*({\bf{r}})\psi({\bf{r}})\psi^*({\bf{r}})\psi({\bf{r}})\right\rangle \\
 =&\sum_{i,j}\left\langle n_in_j\right\rangle+2\sum_{i\neq j}\left\langle n_{i}n_{j}\cos^2(\varphi_i-\varphi_j)\right\rangle\\
=&\,m\left\langle n^{2}\right\rangle+2m(m-1)\left\langle n\right\rangle^{2}\\
=& 2m^{2}\left\langle n\right\rangle^{2},
\end{aligned}
\end{equation}
where $\left\langle n^2\right\rangle=2\left\langle n\right\rangle^2$ for the parametrically driven jet emission as shown in Ref. \onlinecite{Cheng_2017_supp}. %Here we assume CHECK that  since the fluctuations of the jet density is strong such that some angles might have nearly zero occupation.

We compare this with the correlation between forward and backward modes
which is given by
\begin{equation}
\begin{aligned}
&\left\langle \psi^*({\bf{r}})\psi({\bf{r}})\psi^*(-{\bf{r}})\psi(-{\bf{r}})\right\rangle  =\left\langle n_{1}^{2}\right\rangle+\sum_{i,j=2}^{m}\left\langle n_{i}n_{j}^{\prime}\right\rangle\\
 +&\sum_{j=2}^{m}\left\langle n_{1}n_{j}\right\rangle +\sum_{j=2}^{m}\left\langle n_{1}n_{j}^{\prime}\right\rangle\\
=&\left\langle n^{2}\right\rangle+(m-1)^{2}\left\langle n\right\rangle^{2}+2(m-1)\left\langle n\right\rangle^{2}\\
 = &\,(m^{2}+1)\left\langle n\right\rangle^{2}.
\end{aligned}
\end{equation}
We can then write for the real-space correlation functions
\begin{equation}\label{eq:g2}
\begin{aligned}
g^{(2)}(0) & = 2\\
g^{(2)}(\pi) & = 1+\frac{1}{m^{2}}=1+\frac{1}{1+\arctan(R/vt)^{2}/\delta\theta^2}.
\end{aligned}
\end{equation}
In very near field, $m\gg1$, $g_{2}(\pi)\approx 1$, indicating a nearly
full suppression of the $\pi$ peak or equivalently very strong asymmetry.
In the far-field limit
($m\sim1$)
we have $g^{(2)}(\pi)=g^{(2)}(0)\approx 2$,
so that perfect symmetry is restored.

We quantify the asymmetry between peaks at $\phi=0$ and $\phi=\pi$ in real space through
a function
\begin{equation}\label{eq:asym}
\begin{aligned}
\eta_r & =\frac{\left\langle\left[N(\theta)-N(\theta+\pi)\right]^2\right\rangle}{2\left\langle N(\theta)\right\rangle^2}\\
&=\frac{\left\langle\left[n(\theta)-n(\theta+\pi)\right]^2\right\rangle}{2\left\langle n(\theta)\right\rangle^2}\\
%&=\frac{2\pi<\int d\theta \left[n(\theta)^2+n(\theta+\pi)^2-2n(\theta)n(\theta+\pi)\right]>}{2<\int d\theta n(\theta)>^2}\\
&=g^{(2)}(0)-g^{(2)}(\pi)\\
 & =1-\frac{1}{1+\arctan(R/vt)^{2}/\delta\theta^2}.
\end{aligned}
\end{equation}
For momentum space, we similarly define the
analogous function $\eta_k$ using the same expression as for $\eta_r$
but with the real-space population $N(\theta)$ replaced by the momentum-space occupation $N_{k_f}(\theta).$
A fit of the asymmetry function $\eta_r$ by Eq.~\eqref{eq:asym} is given by the solid line (brown) in Fig.~\ref{fig:pipeak}~(c), where the agreement is quite satisfactory.

%The discrepancy at large $t$, as mentioned before, is caused by the neglect of contributions from less relevant jets. This neglect is equivalent to assuming the jet density as a step function, which immediately drops to zero when goes out of its angular width. In fact, the density decrease is smooth. This density tail gives a weak overlap even in large $t$, which effectively adds some 'fractional value' to $m$. This converts our derived abrupt decrease of $\eta$ to a smooth crossover to zero value.

We finally look at
Fig.~\ref{fig:phase}
which addresses the phase coherence of the different modes.
This serves to motivate the analytical model in
Eq.~\ref{eq:jets}.
Plotted in this figure is the
phase of the full GP wavefunction in the far-field configuration, as a function of
position. This is overlaid with a momentum distribution
plot (yellow line) which indicates the real space configuration of the jets in the far field.
``Phase slips" are evident with
varying azimuthal position. Dashed lines have been inserted to
mark these phase discontinuities; one can see that
the phase slips occur somewhere in the empty space between the jets (which each represent a single mode in the far field). Importantly there is phase coherence within
a jet while different modes have uncorrelated phases.
All of this is consistent with the analytical model discussed above and
serves as a validation.


\begin{thebibliography}{28}%
\makeatletter
\providecommand \@ifxundefined [1]{%
 \@ifx{#1\undefined}
}%
\providecommand \@ifnum [1]{%
 \ifnum #1\expandafter \@firstoftwo
 \else \expandafter \@secondoftwo
 \fi
}%
\providecommand \@ifx [1]{%
 \ifx #1\expandafter \@firstoftwo
 \else \expandafter \@secondoftwo
 \fi
}%
\providecommand \natexlab [1]{#1}%
\providecommand \enquote  [1]{``#1''}%
\providecommand \bibnamefont  [1]{#1}%
\providecommand \bibfnamefont [1]{#1}%
\providecommand \citenamefont [1]{#1}%
\providecommand \href@noop [0]{\@secondoftwo}%
\providecommand \href [0]{\begingroup \@sanitize@url \@href}%
\providecommand \@href[1]{\@@startlink{#1}\@@href}%
\providecommand \@@href[1]{\endgroup#1\@@endlink}%
\providecommand \@sanitize@url [0]{\catcode `\\12\catcode `\$12\catcode
  `\&12\catcode `\#12\catcode `\^12\catcode `\_12\catcode `\%12\relax}%
\providecommand \@@startlink[1]{}%
\providecommand \@@endlink[0]{}%
\providecommand \url  [0]{\begingroup\@sanitize@url \@url }%
\providecommand \@url [1]{\endgroup\@href {#1}{\urlprefix }}%
\providecommand \urlprefix  [0]{URL }%
\providecommand \Eprint [0]{\href }%
\providecommand \doibase [0]{http://dx.doi.org/}%
\providecommand \selectlanguage [0]{\@gobble}%
\providecommand \bibinfo  [0]{\@secondoftwo}%
\providecommand \bibfield  [0]{\@secondoftwo}%
\providecommand \translation [1]{[#1]}%
\providecommand \BibitemOpen [0]{}%
\providecommand \bibitemStop [0]{}%
\providecommand \bibitemNoStop [0]{.\EOS\space}%
\providecommand \EOS [0]{\spacefactor3000\relax}%
\providecommand \BibitemShut  [1]{\csname bibitem#1\endcsname}%
\let\auto@bib@innerbib\@empty
%</preamble>
\bibitem [{\citenamefont {Eckardt}(2017)}]{Eckardt}%
  \BibitemOpen
  \bibfield  {author} {\bibinfo {author} {\bibfnamefont {A.}~\bibnamefont
  {Eckardt}},\ }\href {\doibase 10.1103/RevModPhys.89.011004} {\bibfield
  {journal} {\bibinfo  {journal} {Rev. Mod. Phys.}\ }\textbf {\bibinfo {volume}
  {89}},\ \bibinfo {pages} {011004} (\bibinfo {year} {2017})}\BibitemShut
  {NoStop}%
\bibitem [{\citenamefont {Jotzu}\ \emph {et~al.}(2014)\citenamefont {Jotzu}
  \emph {et~al.}}]{Jotzu_2014}%
  \BibitemOpen
  \bibfield  {author} {\bibinfo {author} {\bibfnamefont {G.}~\bibnamefont
  {Jotzu}} \emph {et~al.},\ }\href {\doibase 10.1038/nature13915} {\bibfield
  {journal} {\bibinfo  {journal} {Nature}\ }\textbf {\bibinfo {volume} {515}},\
  \bibinfo {pages} {237} (\bibinfo {year} {2014})}\BibitemShut {NoStop}%
\bibitem [{\citenamefont {Aidelsburger}\ \emph {et~al.}(2015)\citenamefont
  {Aidelsburger} \emph {et~al.}}]{Monika_chern}%
  \BibitemOpen
  \bibfield  {author} {\bibinfo {author} {\bibfnamefont {M.}~\bibnamefont
  {Aidelsburger}} \emph {et~al.},\ }\href {\doibase 10.1038/nphys3171}
  {\bibfield  {journal} {\bibinfo  {journal} {Nature Physics}\ }\textbf
  {\bibinfo {volume} {11}},\ \bibinfo {pages} {162} (\bibinfo {year}
  {2015})}\BibitemShut {NoStop}%
\bibitem [{\citenamefont {Aidelsburger}\ \emph {et~al.}(2011)\citenamefont
  {Aidelsburger} \emph {et~al.}}]{Monika_gauge}%
  \BibitemOpen
  \bibfield  {author} {\bibinfo {author} {\bibfnamefont {M.}~\bibnamefont
  {Aidelsburger}} \emph {et~al.},\ }\href {\doibase
  10.1103/PhysRevLett.107.255301} {\bibfield  {journal} {\bibinfo  {journal}
  {Phys. Rev. Lett.}\ }\textbf {\bibinfo {volume} {107}},\ \bibinfo {pages}
  {255301} (\bibinfo {year} {2011})}\BibitemShut {NoStop}%
\bibitem [{\citenamefont {Zenesini}\ \emph {et~al.}(2009)\citenamefont
  {Zenesini}, \citenamefont {Lignier}, \citenamefont {Ciampini}, \citenamefont
  {Morsch},\ and\ \citenamefont {Arimondo}}]{Zenesini_mott}%
  \BibitemOpen
  \bibfield  {author} {\bibinfo {author} {\bibfnamefont {A.}~\bibnamefont
  {Zenesini}}, \bibinfo {author} {\bibfnamefont {H.}~\bibnamefont {Lignier}},
  \bibinfo {author} {\bibfnamefont {D.}~\bibnamefont {Ciampini}}, \bibinfo
  {author} {\bibfnamefont {O.}~\bibnamefont {Morsch}}, \ and\ \bibinfo {author}
  {\bibfnamefont {E.}~\bibnamefont {Arimondo}},\ }\href {\doibase
  10.1103/PhysRevLett.102.100403} {\bibfield  {journal} {\bibinfo  {journal}
  {Phys. Rev. Lett.}\ }\textbf {\bibinfo {volume} {102}},\ \bibinfo {pages}
  {100403} (\bibinfo {year} {2009})}\BibitemShut {NoStop}%
\bibitem [{\citenamefont {Polkovnikov}\ \emph {et~al.}(2011)\citenamefont
  {Polkovnikov}, \citenamefont {Sengupta}, \citenamefont {Silva},\ and\
  \citenamefont {Vengalattore}}]{Vengalattore}%
  \BibitemOpen
  \bibfield  {author} {\bibinfo {author} {\bibfnamefont {A.}~\bibnamefont
  {Polkovnikov}}, \bibinfo {author} {\bibfnamefont {K.}~\bibnamefont
  {Sengupta}}, \bibinfo {author} {\bibfnamefont {A.}~\bibnamefont {Silva}}, \
  and\ \bibinfo {author} {\bibfnamefont {M.}~\bibnamefont {Vengalattore}},\
  }\href {\doibase 10.1103/RevModPhys.83.863} {\bibfield  {journal} {\bibinfo
  {journal} {Rev. Mod. Phys.}\ }\textbf {\bibinfo {volume} {83}},\ \bibinfo
  {pages} {863} (\bibinfo {year} {2011})}\BibitemShut {NoStop}%
\bibitem [{\citenamefont {Chin}\ \emph {et~al.}(2010)\citenamefont {Chin},
  \citenamefont {Grimm}, \citenamefont {Julienne},\ and\ \citenamefont
  {Tiesinga}}]{ChengFeshbach}%
  \BibitemOpen
  \bibfield  {author} {\bibinfo {author} {\bibfnamefont {C.}~\bibnamefont
  {Chin}}, \bibinfo {author} {\bibfnamefont {R.}~\bibnamefont {Grimm}},
  \bibinfo {author} {\bibfnamefont {P.}~\bibnamefont {Julienne}}, \ and\
  \bibinfo {author} {\bibfnamefont {E.}~\bibnamefont {Tiesinga}},\ }\href
  {\doibase 10.1103/RevModPhys.82.1225} {\bibfield  {journal} {\bibinfo
  {journal} {Rev. Mod. Phys.}\ }\textbf {\bibinfo {volume} {82}},\ \bibinfo
  {pages} {1225} (\bibinfo {year} {2010})}\BibitemShut {NoStop}%
\bibitem [{\citenamefont {Clark}\ \emph {et~al.}(2017)\citenamefont {Clark},
  \citenamefont {Gaj}, \citenamefont {Feng},\ and\ \citenamefont
  {Chin}}]{Cheng_2017}%
  \BibitemOpen
  \bibfield  {author} {\bibinfo {author} {\bibfnamefont {L.~W.}\ \bibnamefont
  {Clark}}, \bibinfo {author} {\bibfnamefont {A.}~\bibnamefont {Gaj}}, \bibinfo
  {author} {\bibfnamefont {L.}~\bibnamefont {Feng}}, \ and\ \bibinfo {author}
  {\bibfnamefont {C.}~\bibnamefont {Chin}},\ }\href {\doibase
  10.1038/nature24272} {\bibfield  {journal} {\bibinfo  {journal} {Nature}\
  }\textbf {\bibinfo {volume} {551}},\ \bibinfo {pages} {356} (\bibinfo {year}
  {2017})}\BibitemShut {NoStop}%
\bibitem [{\citenamefont {Clark}\ \emph {et~al.}(2018)\citenamefont {Clark},
  \citenamefont {Anderson}, \citenamefont {Feng}, \citenamefont {Gaj},
  \citenamefont {Levin},\ and\ \citenamefont {Chin}}]{Logan_gauge}%
  \BibitemOpen
  \bibfield  {author} {\bibinfo {author} {\bibfnamefont {L.~W.}\ \bibnamefont
  {Clark}}, \bibinfo {author} {\bibfnamefont {B.~M.}\ \bibnamefont {Anderson}},
  \bibinfo {author} {\bibfnamefont {L.}~\bibnamefont {Feng}}, \bibinfo {author}
  {\bibfnamefont {A.}~\bibnamefont {Gaj}}, \bibinfo {author} {\bibfnamefont
  {K.}~\bibnamefont {Levin}}, \ and\ \bibinfo {author} {\bibfnamefont
  {C.}~\bibnamefont {Chin}},\ }\href {\doibase 10.1103/PhysRevLett.121.030402}
  {\bibfield  {journal} {\bibinfo  {journal} {Phys. Rev. Lett.}\ }\textbf
  {\bibinfo {volume} {121}},\ \bibinfo {pages} {030402} (\bibinfo {year}
  {2018})}\BibitemShut {NoStop}%
\bibitem [{\citenamefont {Pollack}\ \emph {et~al.}(2009)\citenamefont
  {Pollack}, \citenamefont {Dries}, \citenamefont {Junker}, \citenamefont
  {Chen}, \citenamefont {Corcovilos},\ and\ \citenamefont
  {Hulet}}]{Hulet_2009}%
  \BibitemOpen
  \bibfield  {author} {\bibinfo {author} {\bibfnamefont {S.~E.}\ \bibnamefont
  {Pollack}}, \bibinfo {author} {\bibfnamefont {D.}~\bibnamefont {Dries}},
  \bibinfo {author} {\bibfnamefont {M.}~\bibnamefont {Junker}}, \bibinfo
  {author} {\bibfnamefont {Y.~P.}\ \bibnamefont {Chen}}, \bibinfo {author}
  {\bibfnamefont {T.~A.}\ \bibnamefont {Corcovilos}}, \ and\ \bibinfo {author}
  {\bibfnamefont {R.~G.}\ \bibnamefont {Hulet}},\ }\href {\doibase
  10.1103/PhysRevLett.102.090402} {\bibfield  {journal} {\bibinfo  {journal}
  {Phys. Rev. Lett.}\ }\textbf {\bibinfo {volume} {102}},\ \bibinfo {pages}
  {090402} (\bibinfo {year} {2009})}\BibitemShut {NoStop}%
\bibitem [{\citenamefont {Pollack}\ \emph {et~al.}(2010)\citenamefont
  {Pollack}, \citenamefont {Dries}, \citenamefont {Hulet}, \citenamefont
  {Magalh\~aes}, \citenamefont {Henn}, \citenamefont {Ramos}, \citenamefont
  {Caracanhas},\ and\ \citenamefont {Bagnato}}]{Hulet_2010}%
  \BibitemOpen
  \bibfield  {author} {\bibinfo {author} {\bibfnamefont {S.~E.}\ \bibnamefont
  {Pollack}}, \bibinfo {author} {\bibfnamefont {D.}~\bibnamefont {Dries}},
  \bibinfo {author} {\bibfnamefont {R.~G.}\ \bibnamefont {Hulet}}, \bibinfo
  {author} {\bibfnamefont {K.~M.~F.}\ \bibnamefont {Magalh\~aes}}, \bibinfo
  {author} {\bibfnamefont {E.~A.~L.}\ \bibnamefont {Henn}}, \bibinfo {author}
  {\bibfnamefont {E.~R.~F.}\ \bibnamefont {Ramos}}, \bibinfo {author}
  {\bibfnamefont {M.~A.}\ \bibnamefont {Caracanhas}}, \ and\ \bibinfo {author}
  {\bibfnamefont {V.~S.}\ \bibnamefont {Bagnato}},\ }\href {\doibase
  10.1103/PhysRevA.81.053627} {\bibfield  {journal} {\bibinfo  {journal} {Phys.
  Rev. A}\ }\textbf {\bibinfo {volume} {81}},\ \bibinfo {pages} {053627}
  (\bibinfo {year} {2010})}\BibitemShut {NoStop}%
\bibitem [{\citenamefont {{Tsatsos}}\ \emph {et~al.}(2017)\citenamefont
  {{Tsatsos}}, \citenamefont {{Nguyen}}, \citenamefont {{Lode}}, \citenamefont
  {{Telles}}, \citenamefont {{Luo}}, \citenamefont {{Bagnato}},\ and\
  \citenamefont {{Hulet}}}]{Hulet_granulation}%
  \BibitemOpen
  \bibfield  {author} {\bibinfo {author} {\bibfnamefont {M.~C.}\ \bibnamefont
  {{Tsatsos}}}, \bibinfo {author} {\bibfnamefont {J.~H.~V.}\ \bibnamefont
  {{Nguyen}}}, \bibinfo {author} {\bibfnamefont {A.~U.~J.}\ \bibnamefont
  {{Lode}}}, \bibinfo {author} {\bibfnamefont {G.~D.}\ \bibnamefont
  {{Telles}}}, \bibinfo {author} {\bibfnamefont {D.}~\bibnamefont {{Luo}}},
  \bibinfo {author} {\bibfnamefont {V.~S.}\ \bibnamefont {{Bagnato}}}, \ and\
  \bibinfo {author} {\bibfnamefont {R.~G.}\ \bibnamefont {{Hulet}}},\
  }\href@noop {} {\bibfield  {journal} {\bibinfo  {journal} {ArXiv e-prints}\ }
  (\bibinfo {year} {2017})},\ \Eprint {http://arxiv.org/abs/1707.04055}
  {arXiv:1707.04055} \BibitemShut {NoStop}%
\bibitem [{\citenamefont {{Feng}}\ \emph {et~al.}(2018)\citenamefont {{Feng}},
  \citenamefont {{Hu}}, \citenamefont {{Clark}},\ and\ \citenamefont
  {{Chin}}}]{Lei}%
  \BibitemOpen
  \bibfield  {author} {\bibinfo {author} {\bibfnamefont {L.}~\bibnamefont
  {{Feng}}}, \bibinfo {author} {\bibfnamefont {J.}~\bibnamefont {{Hu}}},
  \bibinfo {author} {\bibfnamefont {L.~W.}\ \bibnamefont {{Clark}}}, \ and\
  \bibinfo {author} {\bibfnamefont {C.}~\bibnamefont {{Chin}}},\ }\href@noop {}
  {\bibfield  {journal} {\bibinfo  {journal} {ArXiv e-prints}\ } (\bibinfo
  {year} {2018})},\ \Eprint {http://arxiv.org/abs/1803.01786}
  {arXiv:1803.01786} \BibitemShut {NoStop}%
\bibitem [{\citenamefont {Milner}(1991)}]{Milner}%
  \BibitemOpen
  \bibfield  {author} {\bibinfo {author} {\bibfnamefont {S.~T.}\ \bibnamefont
  {Milner}},\ }\href {\doibase 10.1017/S0022112091001970} {\bibfield  {journal}
  {\bibinfo  {journal} {Journal of Fluid Mechanics}\ }\textbf {\bibinfo
  {volume} {225}},\ \bibinfo {pages} {81} (\bibinfo {year} {1991})}\BibitemShut
  {NoStop}%
\bibitem [{\citenamefont {Zhang}\ and\ \citenamefont
  {Vi$\tilde{\textrm{n}}$als}(1997)}]{Vinals}%
  \BibitemOpen
  \bibfield  {author} {\bibinfo {author} {\bibfnamefont {W.}~\bibnamefont
  {Zhang}}\ and\ \bibinfo {author} {\bibfnamefont {J.}~\bibnamefont
  {Vi$\tilde{\textrm{n}}$als}},\ }\href {\doibase 10.1017/S0022112096004764}
  {\bibfield  {journal} {\bibinfo  {journal} {Journal of Fluid Mechanics}\
  }\textbf {\bibinfo {volume} {336}},\ \bibinfo {pages} {301} (\bibinfo {year}
  {1997})}\BibitemShut {NoStop}%
\bibitem [{\citenamefont {Staliunas}\ \emph {et~al.}(2002)\citenamefont
  {Staliunas}, \citenamefont {Longhi},\ and\ \citenamefont
  {de~Valc\'arcel}}]{Valcarcel}%
  \BibitemOpen
  \bibfield  {author} {\bibinfo {author} {\bibfnamefont {K.}~\bibnamefont
  {Staliunas}}, \bibinfo {author} {\bibfnamefont {S.}~\bibnamefont {Longhi}}, \
  and\ \bibinfo {author} {\bibfnamefont {G.~J.}\ \bibnamefont
  {de~Valc\'arcel}},\ }\href {\doibase 10.1103/PhysRevLett.89.210406}
  {\bibfield  {journal} {\bibinfo  {journal} {Phys. Rev. Lett.}\ }\textbf
  {\bibinfo {volume} {89}},\ \bibinfo {pages} {210406} (\bibinfo {year}
  {2002})}\BibitemShut {NoStop}%
\bibitem [{\citenamefont {Kagan}\ and\ \citenamefont {Manakova}(2007)}]{Kagan}%
  \BibitemOpen
  \bibfield  {author} {\bibinfo {author} {\bibfnamefont {Y.}~\bibnamefont
  {Kagan}}\ and\ \bibinfo {author} {\bibfnamefont {L.}~\bibnamefont
  {Manakova}},\ }\href {\doibase 10.1016/j.physleta.2006.09.106} {\bibfield
  {journal} {\bibinfo  {journal} {Physics Letters A}\ }\textbf {\bibinfo
  {volume} {361}},\ \bibinfo {pages} {401 } (\bibinfo {year}
  {2007})}\BibitemShut {NoStop}%
\bibitem [{\citenamefont {Nicolin}\ \emph {et~al.}(2007)\citenamefont
  {Nicolin}, \citenamefont {Carretero-Gonz\'alez},\ and\ \citenamefont
  {Kevrekidis}}]{Kevrekidis}%
  \BibitemOpen
  \bibfield  {author} {\bibinfo {author} {\bibfnamefont {A.~I.}\ \bibnamefont
  {Nicolin}}, \bibinfo {author} {\bibfnamefont {R.}~\bibnamefont
  {Carretero-Gonz\'alez}}, \ and\ \bibinfo {author} {\bibfnamefont {P.~G.}\
  \bibnamefont {Kevrekidis}},\ }\href {\doibase 10.1103/PhysRevA.76.063609}
  {\bibfield  {journal} {\bibinfo  {journal} {Phys. Rev. A}\ }\textbf {\bibinfo
  {volume} {76}},\ \bibinfo {pages} {063609} (\bibinfo {year}
  {2007})}\BibitemShut {NoStop}%
\bibitem [{\citenamefont {Bala\ifmmode~\check{z}\else \v{z}\fi{}}\ \emph
  {et~al.}(2014)\citenamefont {Bala\ifmmode~\check{z}\else \v{z}\fi{}},
  \citenamefont {Paun}, \citenamefont {Nicolin}, \citenamefont
  {Balasubramanian},\ and\ \citenamefont {Ramaswamy}}]{Ramaswamy}%
  \BibitemOpen
  \bibfield  {author} {\bibinfo {author} {\bibfnamefont {A.}~\bibnamefont
  {Bala\ifmmode~\check{z}\else \v{z}\fi{}}}, \bibinfo {author} {\bibfnamefont
  {R.}~\bibnamefont {Paun}}, \bibinfo {author} {\bibfnamefont {A.~I.}\
  \bibnamefont {Nicolin}}, \bibinfo {author} {\bibfnamefont {S.}~\bibnamefont
  {Balasubramanian}}, \ and\ \bibinfo {author} {\bibfnamefont {R.}~\bibnamefont
  {Ramaswamy}},\ }\href {\doibase 10.1103/PhysRevA.89.023609} {\bibfield
  {journal} {\bibinfo  {journal} {Phys. Rev. A}\ }\textbf {\bibinfo {volume}
  {89}},\ \bibinfo {pages} {023609} (\bibinfo {year} {2014})}\BibitemShut
  {NoStop}%
\bibitem [{\citenamefont {Engels}\ \emph {et~al.}(2007)\citenamefont {Engels},
  \citenamefont {Atherton},\ and\ \citenamefont {Hoefer}}]{Engels}%
  \BibitemOpen
  \bibfield  {author} {\bibinfo {author} {\bibfnamefont {P.}~\bibnamefont
  {Engels}}, \bibinfo {author} {\bibfnamefont {C.}~\bibnamefont {Atherton}}, \
  and\ \bibinfo {author} {\bibfnamefont {M.~A.}\ \bibnamefont {Hoefer}},\
  }\href {\doibase 10.1103/PhysRevLett.98.095301} {\bibfield  {journal}
  {\bibinfo  {journal} {Phys. Rev. Lett.}\ }\textbf {\bibinfo {volume} {98}},\
  \bibinfo {pages} {095301} (\bibinfo {year} {2007})}\BibitemShut {NoStop}%
\bibitem [{\citenamefont {Inouye}\ \emph {et~al.}(1999)\citenamefont {Inouye},
  \citenamefont {Chikkatur}, \citenamefont {Stamper-Kurn}, \citenamefont
  {Stenger}, \citenamefont {Pritchard},\ and\ \citenamefont
  {Ketterle}}]{Ketterle_grating}%
  \BibitemOpen
  \bibfield  {author} {\bibinfo {author} {\bibfnamefont {S.}~\bibnamefont
  {Inouye}}, \bibinfo {author} {\bibfnamefont {A.~P.}\ \bibnamefont
  {Chikkatur}}, \bibinfo {author} {\bibfnamefont {D.~M.}\ \bibnamefont
  {Stamper-Kurn}}, \bibinfo {author} {\bibfnamefont {J.}~\bibnamefont
  {Stenger}}, \bibinfo {author} {\bibfnamefont {D.~E.}\ \bibnamefont
  {Pritchard}}, \ and\ \bibinfo {author} {\bibfnamefont {W.}~\bibnamefont
  {Ketterle}},\ }\href {\doibase 10.1126/science.285.5427.571} {\bibfield
  {journal} {\bibinfo  {journal} {Science}\ }\textbf {\bibinfo {volume}
  {285}},\ \bibinfo {pages} {571} (\bibinfo {year} {1999})}\BibitemShut
  {NoStop}%
\bibitem [{\citenamefont {Scherpelz}\ \emph {et~al.}(2014)\citenamefont
  {Scherpelz}, \citenamefont {Padavi\ifmmode~\acute{c}\else \'{c}\fi{}},
  \citenamefont {Ran\ifmmode~\mbox{\c{c}}\else \c{c}\fi{}on}, \citenamefont
  {Glatz}, \citenamefont {Aranson},\ and\ \citenamefont {Levin}}]{Kathy}%
  \BibitemOpen
  \bibfield  {author} {\bibinfo {author} {\bibfnamefont {P.}~\bibnamefont
  {Scherpelz}}, \bibinfo {author} {\bibfnamefont {K.}~\bibnamefont
  {Padavi\ifmmode~\acute{c}\else \'{c}\fi{}}}, \bibinfo {author} {\bibfnamefont
  {A.}~\bibnamefont {Ran\ifmmode~\mbox{\c{c}}\else \c{c}\fi{}on}}, \bibinfo
  {author} {\bibfnamefont {A.}~\bibnamefont {Glatz}}, \bibinfo {author}
  {\bibfnamefont {I.~S.}\ \bibnamefont {Aranson}}, \ and\ \bibinfo {author}
  {\bibfnamefont {K.}~\bibnamefont {Levin}},\ }\href {\doibase
  10.1103/PhysRevLett.113.125301} {\bibfield  {journal} {\bibinfo  {journal}
  {Phys. Rev. Lett.}\ }\textbf {\bibinfo {volume} {113}},\ \bibinfo {pages}
  {125301} (\bibinfo {year} {2014})}\BibitemShut {NoStop}%
\bibitem [{\citenamefont {Anderson}\ \emph {et~al.}(2017)\citenamefont
  {Anderson}, \citenamefont {Clark}, \citenamefont {Crawford}, \citenamefont
  {Glatz}, \citenamefont {Aranson}, \citenamefont {Scherpelz}, \citenamefont
  {Feng}, \citenamefont {Chin},\ and\ \citenamefont {Levin}}]{ourKZ}%
  \BibitemOpen
  \bibfield  {author} {\bibinfo {author} {\bibfnamefont {B.~M.}\ \bibnamefont
  {Anderson}}, \bibinfo {author} {\bibfnamefont {L.~W.}\ \bibnamefont {Clark}},
  \bibinfo {author} {\bibfnamefont {J.}~\bibnamefont {Crawford}}, \bibinfo
  {author} {\bibfnamefont {A.}~\bibnamefont {Glatz}}, \bibinfo {author}
  {\bibfnamefont {I.~S.}\ \bibnamefont {Aranson}}, \bibinfo {author}
  {\bibfnamefont {P.}~\bibnamefont {Scherpelz}}, \bibinfo {author}
  {\bibfnamefont {L.}~\bibnamefont {Feng}}, \bibinfo {author} {\bibfnamefont
  {C.}~\bibnamefont {Chin}}, \ and\ \bibinfo {author} {\bibfnamefont
  {K.}~\bibnamefont {Levin}},\ }\href {\doibase 10.1103/PhysRevLett.118.220401}
  {\bibfield  {journal} {\bibinfo  {journal} {Phys. Rev. Lett.}\ }\textbf
  {\bibinfo {volume} {118}},\ \bibinfo {pages} {220401} (\bibinfo {year}
  {2017})}\BibitemShut {NoStop}%
\bibitem [{Note1()}]{Note1}%
  \BibitemOpen
  \bibinfo {note} {This fluctuation term $\psi _s = \varepsilon _r + i
  \varepsilon _i$ is added to the initial ground state wave function $\psi _0$.
  Here we chose the random variables $\varepsilon _r$ and $ \varepsilon _i$ to
  have a Gaussian probability density function centered around zero with
  standard deviation $\sigma = 0.1|\psi _0|$.}\BibitemShut {Stop}%
\bibitem [{Sup()}]{Supple}%
  \BibitemOpen
  \href@noop {} {}\bibinfo {note} {See Supplement.}\BibitemShut {Stop}%
\bibitem [{\citenamefont {{Wu}}\ and\ \citenamefont {{Zhai}}(2018)}]{Hui}%
  \BibitemOpen
  \bibfield  {author} {\bibinfo {author} {\bibfnamefont {Z.}~\bibnamefont
  {{Wu}}}\ and\ \bibinfo {author} {\bibfnamefont {H.}~\bibnamefont {{Zhai}}},\
  }\href@noop {} {\bibfield  {journal} {\bibinfo  {journal} {ArXiv e-prints}\ }
  (\bibinfo {year} {2018})},\ \Eprint {http://arxiv.org/abs/1804.08251}
  {arXiv:1804.08251} \BibitemShut {NoStop}%
\bibitem [{\citenamefont {{Arratia}}(2018)}]{Arratia}%
  \BibitemOpen
  \bibfield  {author} {\bibinfo {author} {\bibfnamefont {M.}~\bibnamefont
  {{Arratia}}},\ }\href@noop {} {\bibfield  {journal} {\bibinfo  {journal}
  {ArXiv e-prints}\ } (\bibinfo {year} {2018})},\ \Eprint
  {http://arxiv.org/abs/1801.05515} {arXiv:1801.05515} \BibitemShut {NoStop}%
\bibitem [{\citenamefont {Ostermann}\ \emph {et~al.}(2016)\citenamefont
  {Ostermann}, \citenamefont {Piazza},\ and\ \citenamefont
  {Ritsch}}]{Ostermann}%
  \BibitemOpen
  \bibfield  {author} {\bibinfo {author} {\bibfnamefont {S.}~\bibnamefont
  {Ostermann}}, \bibinfo {author} {\bibfnamefont {F.}~\bibnamefont {Piazza}}, \
  and\ \bibinfo {author} {\bibfnamefont {H.}~\bibnamefont {Ritsch}},\ }\href
  {\doibase 10.1103/PhysRevX.6.021026} {\bibfield  {journal} {\bibinfo
  {journal} {Phys. Rev. X}\ }\textbf {\bibinfo {volume} {6}},\ \bibinfo {pages}
  {021026} (\bibinfo {year} {2016})}\BibitemShut {NoStop}%
\end{thebibliography}

\begin{thebibliography}{3}%
\makeatletter
\providecommand \@ifxundefined [1]{%
 \@ifx{#1\undefined}
}%
\providecommand \@ifnum [1]{%
 \ifnum #1\expandafter \@firstoftwo
 \else \expandafter \@secondoftwo
 \fi
}%
\providecommand \@ifx [1]{%
 \ifx #1\expandafter \@firstoftwo
 \else \expandafter \@secondoftwo
 \fi
}%
\providecommand \natexlab [1]{#1}%
\providecommand \enquote  [1]{``#1''}%
\providecommand \bibnamefont  [1]{#1}%
\providecommand \bibfnamefont [1]{#1}%
\providecommand \citenamefont [1]{#1}%
\providecommand \href@noop [0]{\@secondoftwo}%
\providecommand \href [0]{\begingroup \@sanitize@url \@href}%
\providecommand \@href[1]{\@@startlink{#1}\@@href}%
\providecommand \@@href[1]{\endgroup#1\@@endlink}%
\providecommand \@sanitize@url [0]{\catcode `\\12\catcode `\$12\catcode
  `\&12\catcode `\#12\catcode `\^12\catcode `\_12\catcode `\%12\relax}%
\providecommand \@@startlink[1]{}%
\providecommand \@@endlink[0]{}%
\providecommand \url  [0]{\begingroup\@sanitize@url \@url }%
\providecommand \@url [1]{\endgroup\@href {#1}{\urlprefix }}%
\providecommand \urlprefix  [0]{URL }%
\providecommand \Eprint [0]{\href }%
\providecommand \doibase [0]{http://dx.doi.org/}%
\providecommand \selectlanguage [0]{\@gobble}%
\providecommand \bibinfo  [0]{\@secondoftwo}%
\providecommand \bibfield  [0]{\@secondoftwo}%
\providecommand \translation [1]{[#1]}%
\providecommand \BibitemOpen [0]{}%
\providecommand \bibitemStop [0]{}%
\providecommand \bibitemNoStop [0]{.\EOS\space}%
\providecommand \EOS [0]{\spacefactor3000\relax}%
\providecommand \BibitemShut  [1]{\csname bibitem#1\endcsname}%
\let\auto@bib@innerbib\@empty
%</preamble>
\bibitem [{\citenamefont {{Arratia}}(2018)}]{Arratia_supp}%
  \BibitemOpen
  \bibfield  {author} {\bibinfo {author} {\bibfnamefont {M.}~\bibnamefont
  {{Arratia}}},\ }\href@noop {} {\bibfield  {journal} {\bibinfo  {journal}
  {ArXiv e-prints}\ } (\bibinfo {year} {2018})},\ \Eprint
  {http://arxiv.org/abs/1801.05515} {arXiv:1801.05515} \BibitemShut {NoStop}%
\bibitem [{\citenamefont {{Wu}}\ and\ \citenamefont {{Zhai}}(2018)}]{Hui_supp}%
  \BibitemOpen
  \bibfield  {author} {\bibinfo {author} {\bibfnamefont {Z.}~\bibnamefont
  {{Wu}}}\ and\ \bibinfo {author} {\bibfnamefont {H.}~\bibnamefont {{Zhai}}},\
  }\href@noop {} {\bibfield  {journal} {\bibinfo  {journal} {ArXiv e-prints}\ }
  (\bibinfo {year} {2018})},\ \Eprint {http://arxiv.org/abs/1804.08251}
  {arXiv:1804.08251} \BibitemShut {NoStop}%
\bibitem [{\citenamefont {Clark}\ \emph {et~al.}(2017)\citenamefont {Clark},
  \citenamefont {Gaj}, \citenamefont {Feng},\ and\ \citenamefont
  {Chin}}]{Cheng_2017_supp}%
  \BibitemOpen
  \bibfield  {author} {\bibinfo {author} {\bibfnamefont {L.~W.}\ \bibnamefont
  {Clark}}, \bibinfo {author} {\bibfnamefont {A.}~\bibnamefont {Gaj}}, \bibinfo
  {author} {\bibfnamefont {L.}~\bibnamefont {Feng}}, \ and\ \bibinfo {author}
  {\bibfnamefont {C.}~\bibnamefont {Chin}},\ }\href {\doibase
  10.1038/nature24272} {\bibfield  {journal} {\bibinfo  {journal} {Nature}\
  }\textbf {\bibinfo {volume} {551}},\ \bibinfo {pages} {356} (\bibinfo {year}
  {2017})}\BibitemShut {NoStop}%
\end{thebibliography}
\end{document}